\documentclass[a4paper,11pt]{article}
\title{Holographic Lifshitz Superconductors with an Axion Field}
\usepackage{amsmath,amssymb,calc}
\usepackage[bulletsep]{collref}
\usepackage{bbm}
\usepackage[pdftex]{graphicx}
\usepackage{verbatim}
\usepackage{braket}
\usepackage{slashed}
\usepackage{epsfig}
\usepackage{graphicx}
\usepackage{array}
\usepackage{caption}
\usepackage{subcaption}

\newcommand\be{\begin{equation}}
\newcommand\ee{\end{equation}}
\newcommand{\bead}{\begin{aligned}}
\newcommand{\eead}{\end{aligned}}

\newcommand{\bea}{\begin{eqnarray}}
\newcommand{\eea}{\end{eqnarray}}

\newcommand{\nn}{\nonumber}

\newcommand{\Tr} {{\textrm{Tr} }}

\def\beq{\begin{equation}}
\def\eeq{\end{equation}}
\def\id{\protect{{1 \kern-.28em {\rm l}}}}

\def\unit{\relax{\rm 1\kern-.26em I}}

\author{Gianni Tallarita\\Centro de Estudios Cient\'{i}ficos (CECs), Casilla 1469, Valdivia, Chile\\ tallarita@cecs.cl}

\begin{document}
\maketitle
\begin{abstract}
We use a Yang-Mills field coupled to an axion as probes of a black hole with arbitrary Lifshitz scaling to investigate, via holography, superconducting phase transitions of the dual theory with a $\braket{p_x+ip_y}$ condensate. In the relativistic case with no axion, this phase is known to be unstable, the stable phase corresponding to condensates of the $\braket{p_x}$ form. We investigate this stability in theories with non-relativistic scaling. Finally we numerically compute the ``Hall" conductivity of these black holes in the non superconducting phase as a function of their Lifshitz scaling.
\end{abstract}
\section{Introduction}

\quad	One of the many uses of the holographic correspondence between a field theory in $d$ dimensions and a $d+1$ dimensional gravitational theory in $AdS$ space \cite{Maldacena:1997re} is in the context of condensed matter theory (CMT). The general field of $AdS/CMT$ has seen numerous advances ranging between diverse topics (see \cite{Hartnoll:2009sz} for a  review), one of which and possibly the most researched is that of holographic superconductivity. The ultimate goal of this approach to superconductivity is to provide precise answers regarding the physics of strongly coupled high critical temperature superconducting materials. The original proposal involves identifying a superconducting phase transition in the dual field theory with a phase transition in the bulk in which a black hole ``wants" to form scalar hair. Since its first conception \cite{Hartnoll:2008vx} there have been many extensions to the basic model. Two of which, and those that concern the present paper, involve extensions to condensates with vectorial order \cite{Gubser:2008wv}, called p-wave superconductors, and non-relativistic, or Lifshitz, holography. A surprising result of \cite{Gubser:2008wv} is that due to the non-linearities of the Yang-Mills action an isotropic ansatz for the gauge field (one with both $dx$ and $dy$ components) in an $AdS$ black hole background leads to solutions which are unstable to perturbations which turn them into an anisotropic (with only a $dx$ component) form. This stability consideration is a major part of the present investigation. Lifshitz holography, born in \cite{Kachru}, involves models with anisotropic scaling between space and time (Lifshitz scaling). Models of this kind have long been of interest in condensed matter regarding tricritical points and phases with modulated order parameters \cite{Diehl} and it is an important question whether the gauge/gravity duality can be successfully applied to them in the context of Lifshitz holographic superconductivity.   This topic has been the subject of numerous studies \cite{Schaposnik:2012cr}\cite{Momeni:2012tw}\cite{Sin:2009wi}\cite{Zhao:2013pva}\cite{Lu:2013tza}\cite{Bry}\cite{Bu:2012zzb}\cite{Hartnoll:2009ns}\cite{Keranen} and many interesting properties of the holographic dual superconductors are found when one abandons a relativistic setting. In particular both the above topics, vector order and Lifshitz scaling, where combined in \cite{Bu:2012zzb} and \cite{Schaposnik:2012cr}  which investigated, both numerically and analytically, finite temperature aspects of the dual theories in the probe limit.
This paper extends this work by including in addition to a vector order parameter and Lifshitz scaling, an axionic field. In the relativistic setting this extra field has many interesting effects on the dual theory. In \cite{Tallarita:2010vu} vortex solutions of the axion system were found which reproduce known characteristics of self-dual Chern-Simons vortices in flat space and \cite{Tallarita:2010uh} found that using an isotropic ansatz for the gauge field (a inostropic ansatz involving one spatial component of the gauge field does not feel the axion coupling), varying the parameter controlling its coupling to the axion one observes changes in order and critical temperatures of phase transitions of the dual field theory. Furthermore, it was shown in \cite{Aprile:2010yb} in a similar but different model, that the axion field introduces non-diagonal components in the conductivity matrix of the dual theory, even in the absence of a magnetic field. It is an interesting question which we address in this paper whether the axion coupling in combination with a non-relativistic scaling, can serve to stabilise the phases with isotropic ansatze for the gauge field. Another important aspect of the axion in this context, and one which we investigate in this paper, is whether the Lifshitz scaling has observable consequences on the previously mentioned ``Hall" conductivity of these black holes. \newline
\indent	 This paper is organised as follows: in section 2 we introduce the gravitational and gauge field system we will work with in the rest of the paper. Here we introduce the axion field, propose an ansatz for the gauge field and discuss the implications in the dual field theory of bulk field behaviour via the holographic dictionary, in section 3 we present the solutions we obtain with this ansatz and demonstrate, through the free energy consideration of section 4, that these solutions are energetically preferred with respect to a Lifshitz black hole with no order parameter. Section 5 is devoted to the stability analysis of the normal and condensed phases and section 6 to the conductivity matrix of the Lifshitz black holes. Finally, in section 7, we draw the conclusions of our investigation.

\section{The System}
We will take as background geometry a general four dimensional Lifshitz black hole with flat horizon topology of the form
\be
ds^2 = L^2\left(-g_z(r)r^{2z}dt^2+\frac{1}{g_z(r)r^2}dr^2+r^2(dx^2+dy^2)\right)
\ee
where
\be
g_z(r)=1-\left(\frac{r_h}{r}\right)^{2z+1},
\ee
and $r_h \leq r < \infty$. In general, black hole solutions with Lifshitz scaling are hard to find, especially those that have analytic expressions for the black hole function $g_z$ and flat horizon topology \cite{Taylor}\cite{Danielsson}\cite{Mann}\cite{Bertoldi}\cite{Bala}\cite{Pang}\cite{Dehghani}\cite{Bravo-Gaete:2013dca}. It is therefore not a priori trivial that this black hole exists. This ansatz creates an excellent toy model in which the basic characteristics of holographic systems with Lifshitz scaling can be appropriately analysed. Importantly, the relativistic case corresponding to $z=1$ reproduces the well known $AdS$ black hole used in \cite{Gubser:2008wv} which, in absence of the axion coupling, serves as a reference point of our calculations.  We point out that this ansatz was used in \cite{Hartnoll:2009ns} for a similar purpose of investigating holographic properties of strange metallic phases. Furthermore, this black hole corresponds to a four dimensional solution found recently in \cite{Bravo-Gaete:2013dca}. \newline
\indent Under the transformation $u=r_h/r$, $0 < u \leq 1 $ we can recast the metric in the form
\be\label{metric}
ds^2=L^2\left(-g_z(u)\frac{r_h^{2z}}{u^{2z}}dt^2+\frac{1}{g_z(u)u^2}du^2+\frac{r_h^2}{u^2}(dx^2+dy^2)\right)
\ee
with 
\be
g_z(u)=1-u^{2z+1}.
\ee
In mass dimensions we have $[t]=-z, [u]=0, [r_h]=1, [x]=[y]=-1$ and for this background $\sqrt{-g}=L^4r_h^{z+2}/u^{z+3}$. The metric is invariant under the rescaling
\be
t\rightarrow \lambda^zt,\quad x\rightarrow \lambda x, \quad u\rightarrow \lambda u
\ee
where the coefficient $z$ is responsible for the non-relativistic invariance, this latter only being valid at $z=1$.
The temperature \cite{Bekenstein:1973ur} of this black hole (from this moment on we will work in units of $L=1$) is
\be
T = \frac{(2z+1)\,r_h^z}{4\pi},
\ee
which sets the temperature of the dual field theory. \newline

On this background we wish to couple an $SU(2)$ Yang-Mills system to an axion field, we make the assumptions that the fields are small and do not backreact on the geometry which will therefore be fixed and uncharged. This corresponds to working in the probe limit. The action we consider takes the form
\be\label{action}
S= \int d^4x \sqrt{-g}\left(-\frac{1}{4}\;\Tr \;F_{\mu\nu}F^{\mu\nu}+\frac{1}{2}\partial_\mu\theta\partial^\mu\theta+\frac{\kappa}{\sqrt{g}}\;\theta\;\Tr(F\wedge F)+V(\theta)\right)
\ee
where
\be
F_{\mu\nu}^a=\partial_\mu A_\nu^a-\partial_\nu A_\mu^a+\epsilon^{abc}A_\mu^bA_\nu^c
\ee
with $\tau^a$ generators of the $SU(2)$ group $[\tau^a,\tau^b]=\epsilon^{abc}\tau^c$ and the metric taking the form described by eq. (\ref{metric}). In our conventions $\epsilon_{tuxy}=+1$. The relevant length scale in the system is the $AdS$ radius $L$ which is assumed large (compared to the string scale) in the semiclassical limit of the holographic conjecture. Given that the axion coupling is a parameter with dimensions of length then physically at the point where $\kappa \approx L$ the probe approximation begins to be unreliable. Working in units of $L$ amounts to requiring $\kappa <1$ in order for the probe approximation to be reliable.\newline

\begin{comment}
An alternative way of thinking about the probe limit is the combined system
\be
S_p = S + S_{grav}
\ee
where
\be
S_{grav} = 
\ee

 as a rescaling of the fields and parameters of the form
\be
F\rightarrow \frac{1}{g}F,\quad \theta\rightarrow\frac{1}{g}\theta,\quad \kappa\rightarrow g\kappa
\ee
and then taking the large $g$ limit.
\end{comment}

 For the gauge field we will pick the ansatz proposed in \cite{Gubser:2008zu}
\be\label{ansatz}
A =\phi(u)\tau^3dt+\omega(u)(\tau^1dx+\tau^2dy).
\ee

\noindent In the relativistic case, there is an important question regarding the stability of this ansatz. It was shown in \cite{Gubser:2008wv} that, in the absence of any axion, this ansatz is unstable towards collapsing into the less isotropic form. This issue is investigated in section 4. The mass dimensions of these fields are $[\phi]=z, [\omega]=1$, also $[\theta]=1, [\kappa]=-1$. \newline

The equations of motion for the gauge and axion fields with the ansatz eq (\ref{ansatz}) reduce to

\be\label{eom1}
\phi''+\frac{(z-1)}{u}\phi'-\frac{2}{r_h^2g_z(u)}\omega^2\phi-\frac{\kappa}{3r_h^{2-z}u^{z-1}}\theta'\omega^2=0
\ee
\be
\omega''+\frac{\partial_u(u^{1-z}g_z(u))}{u^{1-z}g_z(u)}\omega'+\frac{u^{z-1}}{r_h^{2z}g_z^2(u)u^{1-z}}\phi^2\omega-\frac{\omega^3}{r_h^2g_z(u)}+\frac{\kappa}{3u^{1-z}r_h^zg_z(u)}\theta'\phi\omega=0.
\ee
\be\label{eom3}
\theta''+\frac{u^{z+1}}{g_z(u)}\partial_u\left(\frac{g_z(u)}{u^{z+1}}\right)\theta'-\frac{1}{u^2g_z(u)}\frac{\partial V(\theta)}{\partial\theta}+\frac{u^{z+1}}{3g_z(u)r_h^{z+2}}\kappa(\phi\omega^2)'=0.
\ee

Note that the solution for the normal (uncondensed) phase (valid at all $u$) is
\be\label{normal}
\phi = \mu +\rho\log(u),\quad \omega=0 \quad z=2
\ee
\be
\phi = \mu+\rho u^{2-z},\quad \omega=0 \quad z>2.
\ee
and the axion field decouples. Using the horizon condition for $\phi$ we can also set $\mu =0$ and $\mu=(z-2)\rho$ for these solutions. \newline

In order to find phases in which an order parameter appears we solve eqs (\ref{eom1}-\ref{eom3}) using the following boundary behaviour for the fields:
\bea
\phi &=& \phi_1(1-u)+\phi_2(1-u)^2+....\\
\omega&=&\omega_0+\omega_1(1-u)+\omega_2(1-u)^2+...\\
\theta&=&\theta_0+\theta_1(1-u)+\theta_2(1-u)^2+.....
\eea
at the horizon $u=1$. In general $\phi_2,\omega_1...$ etc can be solved in terms of $\phi_1, \omega_0$ and are also functions of the scaling $z$ and $r_h$. 
The boundary $u=0$ behaviours of the gauge field components are
\be\label{bdry1}
\phi = \mu +\rho\log(u)+..., \quad z=2
\ee
\be
\phi = \mu+\rho u^{2-z}+..., \quad z>2.
\ee
and
\be\label{bbdry2}
\omega =\omega_0+ \Omega u^z+....
\ee
According to the gauge/gravity correspondence $\mu$ will be identified with the chemical potential and $\rho$ with the total charge density in the dual theory defined on the boundary. If one is to look for unsourced spontaneous symmetry breaking solutions in the dual theory the constant term $\omega_0$ is required to vanish \cite{Gubser:2008wv}. Then the constant $\Omega$ sets the vacuum expectation value of the order parameter of the phase transition.
The field expansion for $\theta$ depends in general on its potential. If we consider first the case where the potential vanishes, then at the boundary
\be\label{bdry2}
\theta = \theta^b_0 + \frac{\theta_1^b}{2+z}u^{2+z}+...,
\ee
where $ \theta^b_0$ and $\theta_1^b$ are constants,
\begin{comment}
\subsection{Some analytic relations}
We can infer the dependence of boundary quantities analytically by back substitution of the boundary expansions into the equations of motion. Everything can be solved in terms of $\phi_1$ and $\omega_0$ at the horizon,
\be
\phi_2 = \frac{1}{2r_h^4z}\phi_1\left(r_h^4z(z-1)+r_h^2\omega_0-64\kappa^2\omega_0^4\right)
\ee
\be
\omega_1 = \frac{1}{2r_h^2z}\omega_0^3
\ee
\be
\omega_2 = \frac{1}{16r_h^{2(2+z)}z^2}\left(r_h^{2z}\omega_0^3(2r_h^2z^2+3\omega_0^2)+\phi_1^2r_h^2(-r_h^2\omega_0+128\kappa^2\omega_0^3)\right)
\ee
\be
\theta_1 = \frac{8\kappa\phi_1\omega_0^2}{r_h^{2+z}z}
\ee
\be
\theta_2=\frac{2 r_h^{-6-z} \kappa  \phi_1 \omega_0^2 \left(r_h^4 (-4+z) z+3 r_h^2 \omega_0^2-64 \kappa ^2 \omega_0^4\right)}{z^2}.
\ee
In order to get some analytical feel for the behaviour of the functions we match the boundary expansions of the fields and its derivatives at an intermediate point $y$. Then solving the equations
\be
\theta_{r_h}(y)=\theta_b(y), \quad \theta'_{r_h}(y)=\theta'_{b}(y)
\ee
we have
\end{comment}
if instead we consider giving the axion a mass by setting $V(\theta) = \frac{1}{2}m^2\theta^2$, then the axion satisfies the equation of motion of a massive scalar in $AdS_{\tilde{d}+1}$, where $\tilde{d}$ depends on the Lifshitz scaling parameter $z$ as we will shortly show. Consider first the equation of motion of a free massive scalar field in the $AdS_{d+1}$ background without Lifshitz scaling
\be
\frac{1}{\sqrt{-g}}\partial_\mu\left(\sqrt{-g}g^{\mu\nu}\partial_\nu\phi\right)-m^2\phi =0.
\ee
Asymptotically, and for an ansatz which depends only on the $AdS$ variable $u$, this reduces to
\be\label{asym}
u^{d+1}\partial_{u}(u^{-d+1}\partial_u\phi)-m^2\phi=0
\ee
and $\phi$ has boundary behaviour of the form
\be
\phi = c_1 u^{\Delta_+}+c_2 u^{\Delta_-}
\ee
with 
\be
\Delta_{\pm} = \frac{1}{2}\left(d\pm\sqrt{d^2+4m^2}\right),
\ee
and $c_1$, $c_2$ constants. The asymptotic equation for the massive Lifshitz axion becomes
\be
u^{z+3}\partial_u(u^{-z-1}\partial_u\theta)-m^2\theta=0
\ee
which, upon writing $\tilde{d}=z+2$, becomes
\be
u^{\tilde{d}+1}\partial_{u}(u^{-\tilde{d}+1}\partial_u\theta)-m^2\theta=0.
\ee
By comparing to eq(\ref{asym})  we may recognise this as the equation of motion of the massive $d+1$ scalar but for a modified dimension $\tilde{d}+1$. Therefore we may infer the asymptotic behaviour of the field as
\be
\theta = c_1 u^{\Delta_+}+c_2 u^{\Delta_-}
\ee
with 
\be
\Delta_{\pm} = \frac{1}{2}\left(\tilde{d}\pm\sqrt{(\tilde{d})^2+4m^2}\right).  
\ee
We therefore have the possibility of diverging $\Delta_-$ and bound $\Delta_+$ solutions  for the axion field at the $u=0$ boundary. This behaviour is a function of the horizon boundary data. Ensuring that we pick appropriate values for $\theta$ at the horizon we can restrict to solutions where $\theta$ vanishes on the boundary. These conditions are enforced throughout the rest of the paper. In the rest of the paper we consider $m^2>0$ thus ensuring we are always above the Breitenlohner-Freedman bound \cite{Breitenlohner:1982jf}.

\section{Solutions}

We will consider numerical solutions to to eq(\ref{eom1})-(\ref{eom3}). Our numerical method is a shooting method from the horizon to the boundary. The numerical accuracy of this procedure is $O(10^{-8})$. 
\subsection{Vanishing potential}
\begin{figure}
\centering
\begin{subfigure}{.5\textwidth}
  \centering
  \includegraphics[width=.9\linewidth]{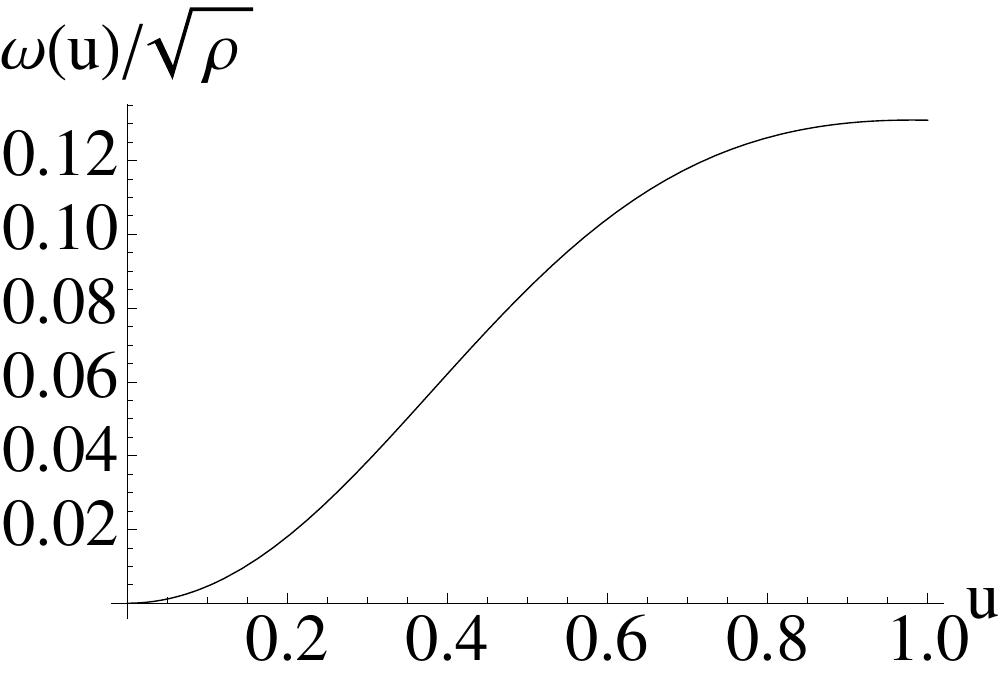}
  \caption{}
  \label{fig:sub1aa}
\end{subfigure}%
\begin{subfigure}{.5\textwidth}
  \centering
  \includegraphics[width=.9\linewidth]{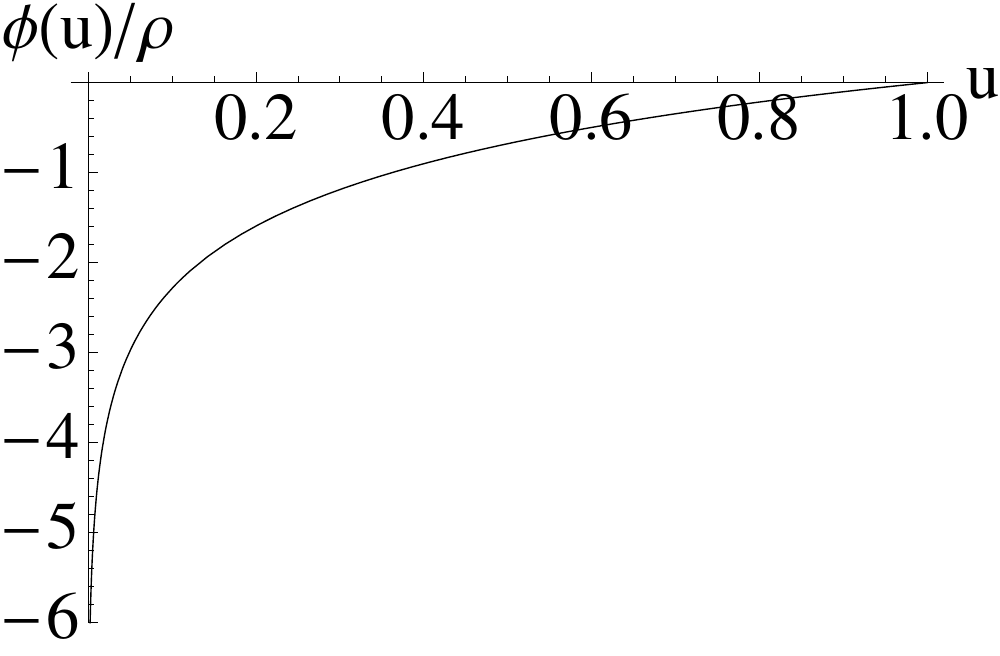}
  \caption{}
  \label{fig:subchi2}
\end{subfigure}
\label{fig:test}
\centering
\begin{subfigure}{.5\textwidth}
  \centering
  \includegraphics[width=.9\linewidth]{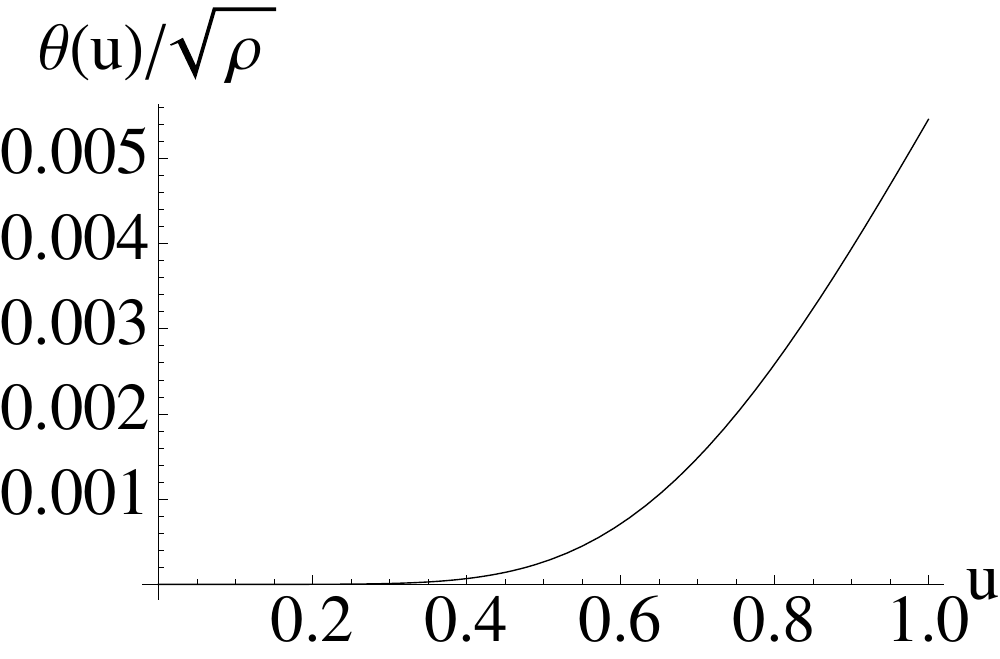}
  \caption{}
  \label{fig:sub22222}
\end{subfigure}
\caption{Field profiles at $z=2$ and $\kappa=1/2$ for $V(\theta)=0$. These field profiles are examples of the general solutions we wish to take for the axion and $\omega$ behaviours. These field profiles show small modifications with $m \neq 0$.}
\label{1}
\end{figure}

We consider first solutions with vanishing potential, \be V(\theta)=0.\ee We are mostly interested in observing the change in the solutions with respect to variations of the axion coupling parameter $\kappa$ and the Lifshitz scaling $z$. We use the asymptotic conditions eq.(\ref{bdry1} - \ref{bbdry2}) for the gauge field and eq.(\ref{bdry2}) for the axion, ensuring, by an appropriate condition on the horizon coefficient, that we have vanishing behaviour at the boundary. As pointed out in \cite{Gubser:2008wv} there are many branches of superconducting solutions characterised by an increasing number of nodes of the $\omega$ function. In general the solution with the least number of nodes (in this case no nodes), is energetically preferred and is the one we shall consider below. The general bulk behaviour of the fields can be seen in Figure \ref{1}.

With $\kappa=0$ the axion field decouples. This case corresponds to the case considered in \cite{Gubser:2008wv} but with Lifshitz scaling where, at least for $z=1$, we expect the ansatz to be unstable. In Figure (\ref{2aaa}) we show the effects of the Lifshitz scaling on the critical temperature of the phase transitions both with and without an axion coupling. We find that raising $z$ we observe a shift in the critical temperature. As the non-relativistic scaling is increased, the shift of critical temperatures is towards zero. A similar behaviour is seen in \cite{Lu:2013tza} for a $\braket{p_x}$ Lifshitz holographic superconductor. When the axion field is coupled to the system we see that its effect on the critical temperatures is small. However, as $\kappa$ is increased we find that the critical temperature decreases. Figures (\ref{fig3aaaa}) and (\ref{fig3bbbb}) show the corresponding profiles of the order parameters for the dual theories, we observe that below the critical temperature the order parameter is non-zero as expected from the phase transition. Close to the critical temperature the condensates fits well to a 
\be
\Omega \approx \sqrt{1-\frac{T}{T_c}}
\ee
behaviour, as expected in the mean field limit.

\begin{figure}
\centering
\begin{subfigure}{.5\textwidth}
  \centering
  \begin{tabular}{| l |c| c | r| }
    \hline
   $z$&  $T_c/\rho$  \\ \hline\hline
    1&0.0653978    \\ \hline
    2&0.0440729   \\ \hline
    3&0.0349058   \\ \hline
    4& 0.0294475 \\ \hline 
  \end{tabular}
  \caption{$\kappa=0$.}
  \label{fig4a}
\end{subfigure}%
\begin{subfigure}{.5\textwidth}
  \centering
   \begin{tabular}{| l | c | r| }
    \hline
   $z$ & $T_c/\rho$ \\ \hline\hline
    1 & 0.0653976 \\ \hline
     2 & 0.0440727 \\ \hline
     3 & 0.0349056\\ \hline
     4 &0.0294474 \\ \hline 
 %   1 & 8 & 0.114362 \\
    %\hline
  \end{tabular}
  \caption{$\kappa=0.8$.}
  \label{fig:subchi2}
\end{subfigure}
\caption{Critical temperatures for $\kappa=0,0.8$ and varying $z$. In (a) we show the effects of varying $z$ in a pure Lifshitz setting with no axion field. In (b) we show the same results with an axion field turned on. $\kappa$ is measured in units of $L$.}
\label{2aaa}
\end{figure}

\subsection{Massive axion}

We now switch on a massive potential for the axion field, \be\label{mass} V(\theta)=\frac{1}{2}m^2\theta^2.\ee Given that the axion decouples if $\kappa=0$ we will restrict here to the case where this doesn't vanish. Reliability of the probe limit restrict us to $mL<1$. The field profile solutions show little variations to those shown in Figure (\ref{1}), owing to the small values of $m$ considered in the probe limit. The results of varying $m$ on the critical temperature of the phase transitions are shown in Figure (\ref{figmassive}). We see that for all Lifshitz scalings considered the effect of raising the axion mass is to raise the critical temperature. This effect is in the opposite direction to that of raising $z$, yet it is too small to observe a critical temperature which is higher at a greater $z$, at least if one remains in the probe limit. In this limit we find that the effects of $m$ are small on the field profiles.
\begin{figure}
\centering
\begin{subfigure}{.5\textwidth}
  \centering
  \includegraphics[width=.9\linewidth]{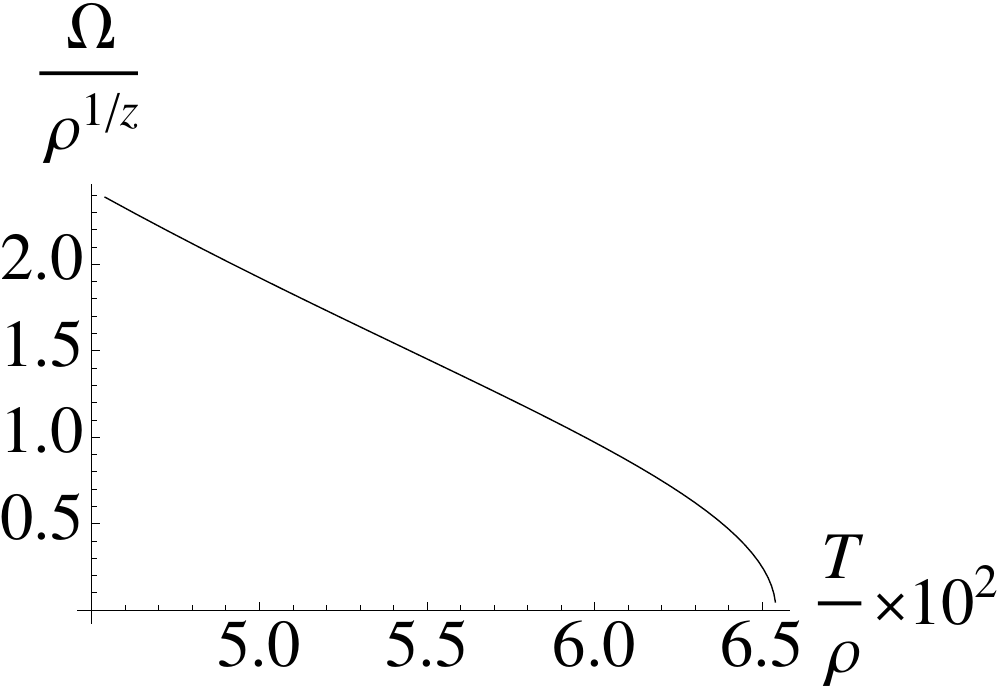}
  \caption{Condensate at $z=1$, $\kappa=0$.}
  \label{fig3aaaa}
\end{subfigure}%
\begin{subfigure}{.5\textwidth}
  \centering
  \includegraphics[width=.9\linewidth]{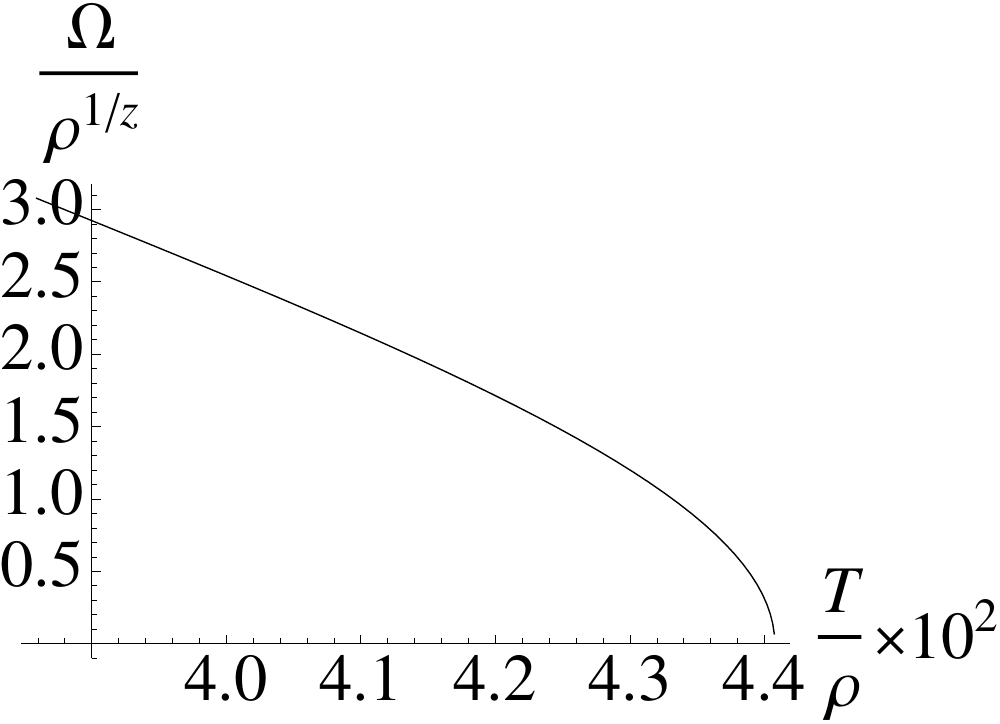}
  \caption{Condensate at $z=2$, $\kappa=0$.}
  \label{fig3bbbb}
\end{subfigure}
\centering

\caption{Plots (a) and (b) show the order parameters at different values of the Lifshitz scaling for $\kappa=0$. Note the shift in the critical temperature. Similar profiles are observed by switching on $\kappa$ and $m$.}
\label{fig1}
\end{figure}

\begin{figure}
\centering
\begin{subfigure}{.5\textwidth}
  \centering
  \begin{tabular}{| l |c| c | r| }
    \hline
  $m$ & $T_c/\rho$ \\ \hline\hline
      0.2 & 0.0440731 \\ \hline
     0.4 &0.0440742\\ \hline
     0.6 &0.0440761\\ \hline
     0.8 &0.044079\\ \hline 
  \end{tabular}
  \caption{$z=2$}
  \label{fig4a}
\end{subfigure}%
\begin{subfigure}{.5\textwidth}
  \centering
   \begin{tabular}{| l | c| c | r| }
    \hline
 $m$ & $T_c/\rho$ \\ \hline\hline
 0.2 &0.0349058 \\ \hline
 0.4 &0.0349063\\ \hline
0.6&0.0349071\\ \hline
 0.8&0.0349083\\ \hline 
    %1 & 8 & 0.114362 \\
    %\hline
  \end{tabular}
  \caption{$z=3$}
  \label{fig4b}
\end{subfigure}
\\
\begin{subfigure}{.5\textwidth}
  \centering
   \begin{tabular}{| l | c| c | r| }
    \hline
 $m$ & $T_c/\rho$ \\ \hline\hline
 0.2 &0.0294475\\ \hline
 0.4 &0.0294478\\ \hline
0.6&0.0294482\\ \hline
 0.8&0.0294489\\ \hline 
    %1 & 8 & 0.114362 \\
    %\hline
  \end{tabular}
  \caption{$z=4$}
  \label{fig4c}
\end{subfigure}
\caption{The variation of critical temperatures of the phase transition with respect to variations of the mass of the axion field at various Lifshitz scalings. These values correspond to $\kappa=0.8$. }
\label{figmassive}
\end{figure}

\section{Free energy}

In this section we wish to compute the difference in free energy densitites between the superconducting and normal phases as a function of Lifshitz scaling, axion coupling and mass of the axion field. If this difference is negative then the superconducting phase is preferred. 
The holographic dictionary relates the free energy density $\mathcal{F}$ of the dual field theory to the Euclidean on-shell action in the bulk according to
\be
\mathcal{F}=T S_E\big|_{\text{on shell}}.
\ee
Substituting our ansatz into the action and performing a Wick rotation one obtains, after an integration by parts
%\be
%S = \int d^4x\;\;-\frac{u^{1-z}r_h^z}{g_z(u)}(\omega')^2+\frac{1}{2}\frac{r_h^{2-z}u^{z-1}}{g_{z}(u)^2}(\phi')^2+\frac{u^{z-1}r_h^{-z}}{g_z(u)}(\phi\omega)^2-\frac{1}{2}u^{1-z}r_h^{z-2}\omega^4\nonumber\ee\be-\frac{1}{2}\partial_u\left(\frac{r_h^{z+2}}{u^{z+1}g_z(u)}\theta'\right)\theta-\frac{1}{3}\kappa\theta(\phi\omega^2)'+\frac{r_h^{z+2}}{u^{z+3}}V(\theta)+\frac{V_2}{T}\frac{1}{2}\left(\sqrt{g}g^{uu}\theta\theta'\right)\big|_{bdry}
%\ee
%once this is euclideanised $\phi\rightarrow i\phi$ and put on-shell the final form is
\be
S = \frac{V_{s}}{T}\bigg\{\int du \left[\frac{1}{2}r_h^{z-2}u^{1-z}\omega^4+\frac{u^{z-1}\omega^2\phi^2}{r_h^zg_z(u)}-\frac{i}{2}\kappa\theta'\phi\omega^2+\frac{r_h^{z+2}}{u^{z+3}}\left(-\theta\frac{\partial V}{\partial\theta}+V(\theta)\right)\right]\nn\ee\be
-\left(u^{1-z}r_h^zg_z(u)\omega'\omega+\frac{1}{2}\frac{r_h^{2-z}u^{z-1}}{g_z(u)^2}\phi'\phi-\frac{1}{2}\frac{r_h^{z+2}g_z(u)}{u^{z+1}}\theta'\theta\right)\big|_{bdry}\bigg\},
\ee
where $V_s$ is the volume (area) of the $x,y$ plane.
Working at fixed charge means we must add to this the term \cite{Skenderis}
\be
S_t = -\frac{1}{2}\int dtd^2x \sqrt{-g}A_\mu F^{u\mu}\big|_{u=0}
\ee
which evaluates to 
\be
S_t = -\frac{V_s}{T}\frac{u^{z-1}r_h^{2-z}}{2}\phi'\phi\big|_{u=0}
\ee
once Euclideanised.
Using the asymptotic behaviour of our solutions  we may reduce the above to 
\be
\mathcal{F} =\int du \left[\frac{1}{2}r_h^{z-2}u^{1-z}\omega^4+\frac{u^{z-1}\omega^2\phi^2}{r_h^zg_z(u)}-\frac{r_h^{z+2}}{u^{z+3}}\left(\theta\frac{\partial V}{\partial\theta}-V\right)\right]
-\frac{r_h^{2-z}u^{z-1}}{g_z(u)^2}\phi'\phi\big|_{bdry}
\ee
where the boundary expression for $\phi$ depends on the choice of $z$. Note that we deliberately excluded the term proportional to $i\kappa$ which does not contribute to the energy of the dual theory (this was previously observed in \cite{Tallarita:2010uh}). \newline

\begin{figure}
\centering
\begin{subfigure}{.5\textwidth}
  \centering
  \includegraphics[width=.9\linewidth]{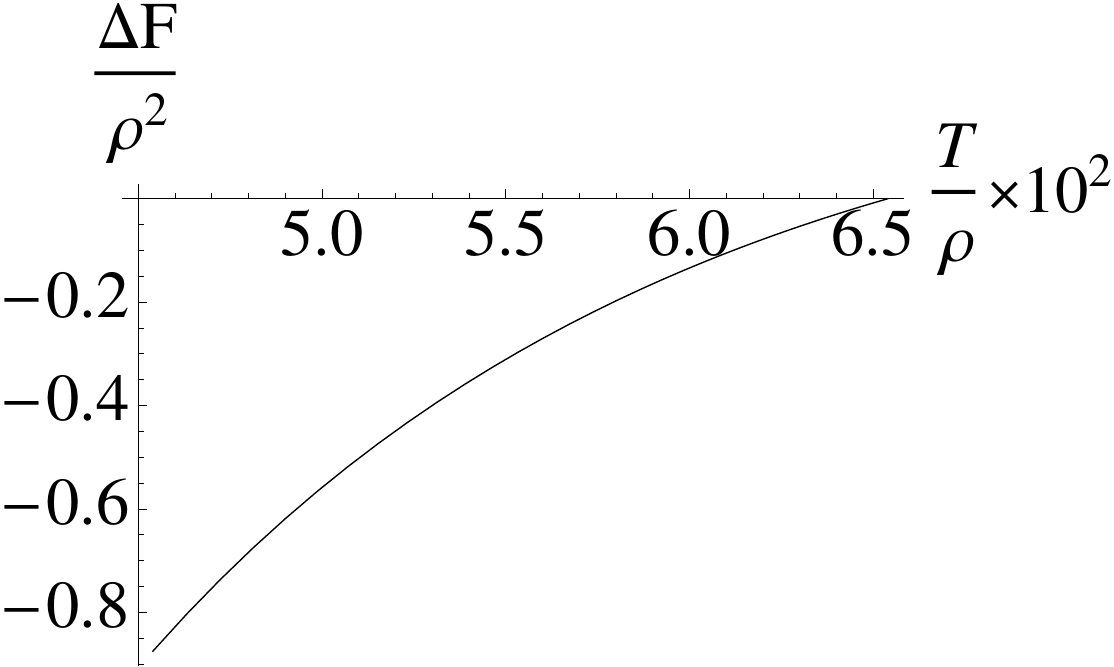}
  \caption{Free energy at $z=1$, $\kappa=0$.}
  \label{fig3c}
\end{subfigure}%
\begin{subfigure}{.5\textwidth}
  \centering
  \includegraphics[width=.9\linewidth]{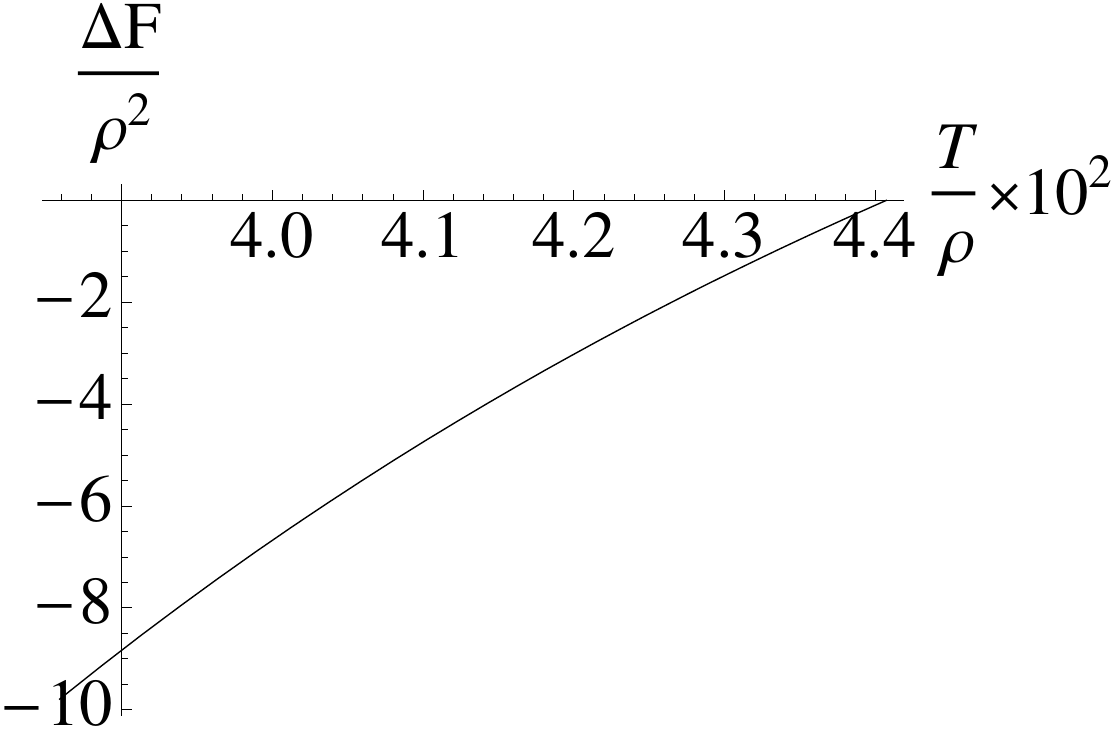}
  \caption{Free energy at $z=2$, $\kappa=0$.}
  \label{fig3d}
\end{subfigure}
\caption{Plots (a) and (b) show the difference in free energy densities between the normal and condensed phases. The plots correspond to $m=0$.}
\end{figure}

Let us consider first the case of vanishing potential and $z=2$. From the last expression substituting the boundary behaviour (\ref{bdry1})-(\ref{bdry2}) we obtain
\be
\mathcal{F} = -\rho\mu+\int_{u=1}^{u=0} du \left(\frac{1}{2u}\omega^4+\frac{u\;\omega^2\phi^2}{r_h^2g_z(u)}\right)-\rho^2\log(u)|_{u=\epsilon}.
\ee
The last term is divergent, this divergence is expected from the behaviour of the gauge field at the boundary Figure (\ref{1}). However we may remove the divergence by considering the difference between the condensed and uncondensed phases, the latter only contributes this divergence using the normal solution eq.(\ref{normal}). Therefore
\be
\Delta \mathcal{F}=\mathcal{F}-\mathcal{F}_n =  -\rho\mu+\int_{u=1}^{u=0} du \left(\frac{1}{2u}\omega^4+\frac{u\;\omega^2\phi^2}{r_h^2g_z(u)}\right).
\ee
When $\Delta\mathcal{F}$ is negative the condensed phase is preferred. Numerical solutions for $z=1$ and the more interesting $z=2$ case are shown in Figures (\ref{fig3c})-(\ref{fig3d}). In this case we find that the free energy difference is negative below the critical temperature, indicating that the condensing phase is preferred. \newline

\begin{figure}
\centering
\begin{subfigure}{.5\textwidth}
  \centering
  \includegraphics[width=.9\linewidth]{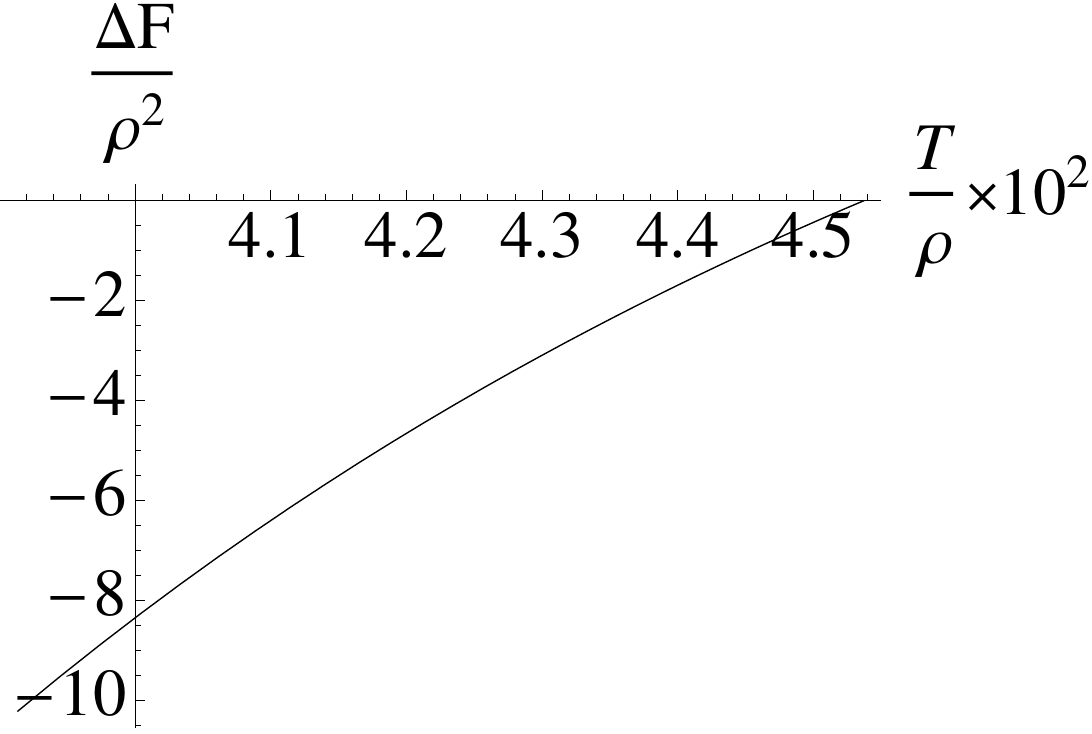}
  \caption{}
  \label{fig:sub1aa}
\end{subfigure}%
\begin{subfigure}{.5\textwidth}
  \centering
  \includegraphics[width=.9\linewidth]{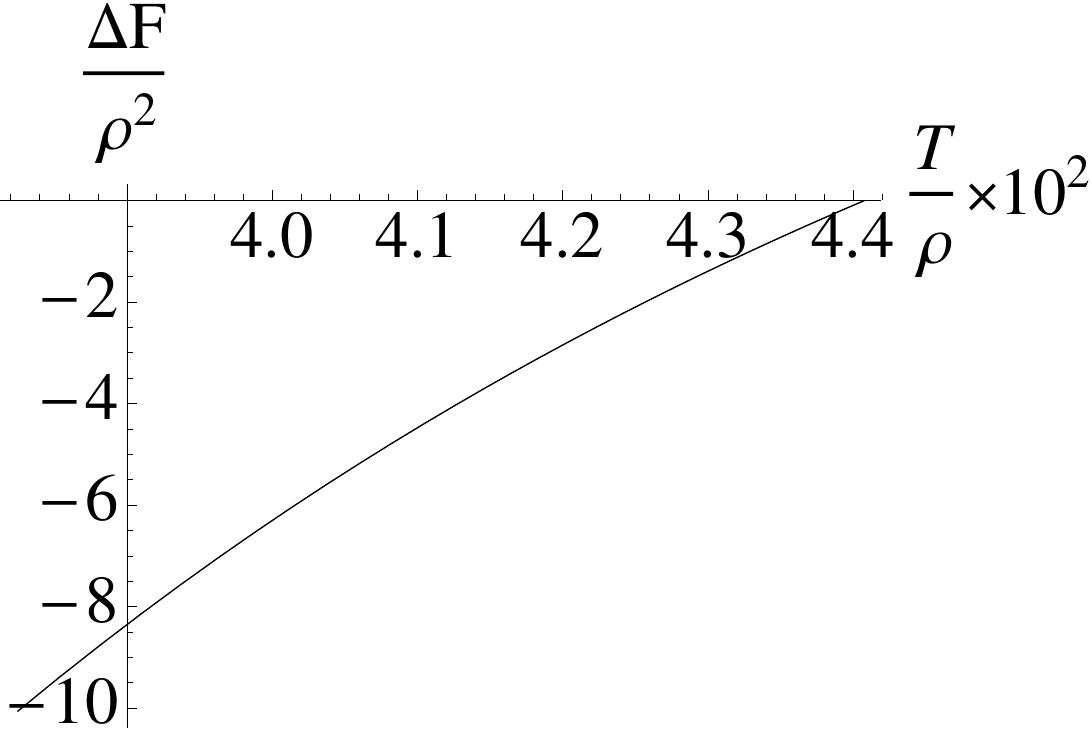}
  \caption{}
  \label{fig:subchi2}
\end{subfigure}
\caption{Difference in free energy densities between the normal and the superconducting phases at $z=2$, $\kappa=0.8$. In (a) we set $m=1$ whilst (b) has $m=0$.}
\label{fig5}
\end{figure}

Now let us switch on the massive potential eq(\ref{mass}) at $z=2$. Then we have
\be
\Delta\mathcal{F} = -\rho\mu+\int_{u=1}^{u=0} du \left(\frac{1}{2u}\omega^4+\frac{u\;\omega^2\phi^2}{r_h^2g_z(u)}-\frac{r_h^{z+2}}{u^{z+3}}m^2\theta^2\right).
\ee
This case is shown in Figure (\ref{fig5}). Given that for our choice of boundary condition, with the axion field vanishing at the boundary, the solution for the axion is small compared to the gauge field (see Figure (\ref{1})) its contribution to the free energy is also small and the main observable difference is the shift in the critical temperature as we expected from Figure (\ref{figmassive}). This plot differs from Figure (\ref{fig3d}) by the value of $\kappa$ thus confirming that this has no large effect in the free energy as we expected from the analytical result. Similar results are obtained for $z=3$ and $z=4$.

% If we pick the $u$ integration from the horizon to the boundary then we see that 
%\be
%\mathcal{F} = -u^{z-1}r_h^{2-z}\phi'\phi\big|_{u=0}+\frac{1}{2}\frac{r_h^{z+2}}{u^{z+1}}\theta'\theta\big|_{u=0}+\int_{u=1}^{u=0} du \left(\frac{1}{2}r_h^{z-2}u^{1-z}\omega^4+\frac{u^{z-1}\omega^2\phi^2}{r_h^zg_z(u)}\right)
%\ee

 \section{Stability Calculation}
 
  It is known \cite{Gubser:2008wv} that in the case of $\kappa=0$ and $z=1$, the relativistic system without an axion, the most stable ansatz is not eq(\ref{ansatz}) but the less isotropic $\braket{p_x}$ form.  This section is devoted to analysing the stability of ansatz eq(\ref{ansatz}) in the presence of both Lifshitz scaling and an axion term.
 We wish to see here whether a non relativistic Lifshitz scaling or an axion coupling can serve to stabilise the isotropic ansatz for the gauge field.\newline
 
 Consider perturbing our ansatz (\ref{ansatz}) so that $A\rightarrow A+a$ where
 \be
 a = e^{-iw t}a_1(\tau^1dx-\tau^2dy)+e^{-iw t}a_2(\tau^2dx+\tau^1dy).
 \ee
These perturbations are the relevant perturbations in order to investigate the stability in the $x, y$ directions. Then the equations of motion of the perturbations at linearized level become
 %\be
 %\frac{1}{2}\partial_m\left[\sqrt{g}g^{\mu m}g^{\nu\tau}(f_{\mu\nu}^d+\epsilon^{dbc}(A_\mu^ba_\nu^c+A_\nu^ca_\mu^b))\right]-2\kappa\epsilon_{\mu\nu\rho\tau}\partial_\rho\left[\theta(f_{\mu\nu}^d+\epsilon^{dbc}(A_\mu^ba_\nu^c+A_\nu^ca_\mu^b)))\right]\ee\be+\frac{1}{2}\sqrt{g}g^{\mu m}g^{\nu\tau}\epsilon^{adb}\left[A_m^b(f_{\mu\nu}^a+\epsilon^{aef}(A_\mu^ea_\nu^f+A_\nu^fa_\mu^e))+a_m^bF_{\mu\nu}^a\right]\ee\be
 %+2\kappa\theta\epsilon_{\mu\nu\tau\rho}\epsilon^{adb}\left[A_\rho^b(f_{\mu\nu}^a+\epsilon^{aef}(A_\mu^ea_\nu^f+A_\nu^fa_\mu^e))+a_\rho^bF_{\mu\nu}^a\right]=0
% \ee
% where
 %\be
 %f_{\mu\nu}^a=\partial_\mu a_\nu^a-\partial_\nu a_\mu^a +\epsilon^{abc}a_\mu^b a_\nu^c.
 %\ee
 \bea
 \bigg(\frac{1}{2}r_h^z(gu^{1-z})\frac{\partial^2}{\partial u^2}+\frac{1}{2}r_h^z\partial_u(gu^{1-z})\frac{\partial}{\partial u}+\frac{1}{2}r_h^{z-2}u^{1-z}\omega^2+\frac{1}{2}r_h^{-z}\frac{u^{z-1}}{g}\phi^2-4\kappa\theta'\phi\bigg)a_1 \nonumber\eea
 \be
 +\left(-\frac{1}{r_h^z}\frac{u^{z-1}}{g}i w\phi+4i\kappa w\theta'\right)a_2=0
 \ee
 and
  \be
 \left(\frac{1}{2}r_h^z(gu^{1-z})\frac{\partial^2}{\partial u^2}+\frac{1}{2}r_h^z\partial_u(gu^{1-z})\frac{\partial}{\partial u}+\frac{1}{2}r_h^{z-2}u^{1-z}\omega^2+\frac{1}{2}r_h^{-z}\frac{u^{z-1}}{g}\phi^2-4\kappa\theta'\phi\right)a_2
 \nonumber\ee
  \be
 -\left(-\frac{1}{r_h^z}\frac{u^{z-1}}{g}i w\phi+4i\kappa w \theta'\right)a_1=0.
 \ee
 We wish to solve these coupled equations with infalling boundary conditions for $a_1$ and $a_2$ at the horizon, which are 
 \be
 a_{x/y} \approx \left(1-u\right)^{-\frac{i w}{(2z+1)r_h^z}}\left(a^0_{x/y}+a^{1}_{x/y}(1-u)+....\right)
 \ee
  and demand that they vanish on the boundary. These conditions are satisfied for certain quasinormal frequencies $w$ only. Frequencies with positive imaginary part correspond, according to our choice of perturbation, to unstable modes of the system. In Figures (6) and (7) we summarize our findings with respect to variations of the Lifshitz scaling and the axion coupling. 
  
   \begin{figure}
\centering
\begin{subfigure}{.5\textwidth}
  \centering
  \begin{tabular}{| l |c| c | r| }
    \hline
    $T/\rho$ & $z$ & $Im(w)/\rho$ \\ \hline\hline
    0.0795775 & 1 &   -0.0381984  \\ \hline
    0.0663146 & 2 &  -0.0974759 \\ \hline
    0.0464202 & 3 &-0.0576729  \\ \hline
    0.0358099 & 4 &-0.0350438\\ \hline 
  \end{tabular}
  \caption{$T > T_c$}
  \label{fig4a}
\end{subfigure}%
\begin{subfigure}{.5\textwidth}
  \centering
   \begin{tabular}{| l | c| c | r| }
    \hline
 $T/\rho$ & $z$ & $Im(w)/\rho$ \\ \hline\hline
  0.0530516 & 1  &0.0223665 \\ \hline
 0.0397887 & 2 &0.0120235\\ \hline
  0.0278521 & 3&0.019798 \\ \hline
 0.0255785 & 4&0.0143312\\ \hline 
    %1 & 8 & 0.114362 \\
    %\hline
  \end{tabular}
  \caption{$T<T_c$}
  \label{fig4b}
\end{subfigure}
\caption{Imaginary parts of quasinormal frequencies at varying values of the Lifshitz scaling $z$ and the axion coupling $\kappa$ for the normal phase with $w=0$ above and below the critical temperature. Here we have $\kappa=0$.}
\label{fig:test}
\end{figure}
 \begin{figure}
\centering
\begin{subfigure}{0.5\textwidth}
  \centering
  \begin{tabular}{| l |c| c | r| }
    \hline
   T/$\rho$ &   $z$ & $Im(w)/\rho$ \\ \hline\hline
  0.0477465 &  1 &   0.00698173  \\ \hline
  0.0426186  & 2 &  0.00255119 \\ \hline
  0.0342888 & 3 & 0.00133075 \\ \hline
   0.0291224  & 4 & 0.000817981\\ \hline 
  \end{tabular}
  \caption{$\kappa=0$}
  \label{fig4a}
\end{subfigure}%
\begin{subfigure}{.5\textwidth}
  \centering
   \begin{tabular}{| l | c| c | r| }
    \hline
  $\kappa$ & $Im(w)/\rho$ \\ \hline\hline
  0.2 & 0.00767316 \\ \hline
   0.4 & 0.00963637\\ \hline
   0.6 &0.0125594 \\ \hline
   1 & 0.0194076\\ \hline 
    %1 & 8 & 0.114362 \\
    %\hline
  \end{tabular}
  \caption{$z=1$, $T/\rho = 0.0598$}
  \label{figstab}
\end{subfigure}\\

\begin{subfigure}{.5\textwidth}
  \centering
  \begin{tabular}{| l |c| c | r| }
    \hline
 $m$ & $Im(w)/\rho$ \\ \hline\hline
   0.4  &   0.0128123  \\ \hline
  0.8  &  0.0124561 \\ \hline
  \end{tabular}
  \caption{$\kappa=0.8$,\; $z=2$, $T/\rho = 0.0426$}
  \label{figstaba}
\end{subfigure}%
\begin{subfigure}{.5\textwidth}
  \centering
   \begin{tabular}{| l | c| c | r| }
    \hline
   $m$ & $Im(w)/\rho$ \\ \hline\hline
 0.4  &   0.0112293  \\ \hline
0.8  &  0.0110764 \\ \hline
  \end{tabular}
  \caption{$\kappa=0.8$,\; $z=4$, $T/\rho = 0.0291$}
  \label{figstab}
\end{subfigure}
\\
\caption{Imaginary parts of quasinormal frequencies at varying values of the Lifshitz scaling $z$ and the axion coupling $\kappa$ at $T<T_c$ for the superconducting ansatz}
\label{figstabat}
\end{figure}
Let us begin by analysing the stability of the normal phase. We see from Figure (6) that for all Lifshitz scalings the black holes in the normal phase (with no hair $\omega=0$) are stable above a critical temperature and unstable below it. The system finds it energetically preferrable to switch on hair below a critical temperature, as we expected from the free energy considerations of section 4. 
In Figure (\ref{figstabat}) we present the analysis of the stability of the condensed phase for various combinations of the couplings. Within the parameter range investigated above, and for all temperatures $T<T_c$ (far enough from $T=0$ where the probe limit is unreliable) we find no evidence that either the axion coupling or a non-relativistic Lifshitz scaling can serve to stabilize the isotropic ansatz. Combinations of diverse values of both parameters were also investigated and no stable mode was found. Hence, as per the case $\kappa=0$ and $z=1$ the solution obtained from the ansatz (\ref{ansatz}) is unstable.
 \section{``Hall" Conductivity of the normal phase}
 
In this section we will consider electromagnetic perturbations of our system described by the action (\ref{action}). Given that the ansatz giving rise to the superconducting phase is unstable (at least within the parameter range explored in this paper) we will not focus on perturbations of this phase but rather of the normal phase with $\omega=0$, which is certainly stable above the critical temperature. As is well known \cite{Herzog:2007ij} perturbations of the normal phase at $z=1$ without any axion coupling lead to a diagonal constant conductivity matrix 
\be
\sigma_{xx}= \sigma_{yy}=1.
\ee

As shown in \cite{Aprile:2010yb} in the context of a similar model the axion coupling introduces non-diagonal, or ``Hall", components of the conductivity matrix in the dual theory, even in the superconducting phase. These components are observed in the absence of any external magnetic field and were therefore interpreted as anomalous or topological. We wish to investigate both $\sigma_{xx}$ and $\sigma_{xy}$ components of the conductivity matrix of the uncondensed phase for varying values of the Lifshitz scaling.   In the normal phase perturbations of the form
\be
a = e^{-i w t}\left((a^t_1\tau_1+a^t_2\tau^2)dt+a_x\tau^3dx+a_y\tau^3dy\right)
\ee
do not mix the $a_{x/y}$ with the $a^t$. Recall also that in this phase the axion field effectively decouples from the system, this does not mean however that perturbations of this phase shouldn't feel the presence of the axion's interaction to the gauge field. The resulting linearised equations of motion for the perturbations are 
\be
\frac{u^{-1+z} (r_h)^{-z} w^2 a_x}{2g}+\frac{1}{2} r_h^{z}\left(u^{1-z} g\right)'(a_x)'+12 i  w \kappa  a_y \theta '+\frac{1}{2} r_h^{z} u^{1-z} g (a_x)''=0
\ee
\be
\frac{u^{-1+z} (r_h)^{-z} w^2 a_y}{2g}+\frac{1}{2} r_h^{z}\left(u^{1-z} g\right)'(a_y)'-12 i  w \kappa  a_x \theta '+\frac{1}{2} r_h^{z} u^{1-z} g (a_y)''=0
\ee
We see that the effect of the axion term is to couple the perturbation modes. The equation for the axion field takes the form
\be
\theta''+\frac{u^{z+1}}{g_z(u)}\partial_u\left(\frac{g_z(u)}{u^{z+1}}\right)\theta'-\frac{1}{u^2g_z(u)}\frac{\partial V(\theta)}{\partial\theta}=0.
\ee
We switch on a small mass so as to avoid solutions in which $\theta = const$ as in this case the mixing term in the perturbation equations disappears and one cannot observe any non-diagonal components of the conductivity matrix. To proceed we decouple the system of equations using
\be
A_\pm = \frac{1}{2}(a_x\pm ia_y)
\ee
then, for these combinations we obtain the following decoupled equations
\be
\frac{u^{-1+z} (r_h)^{-z} w^2 A_+}{2g}+\frac{1}{2} r_h^{z}\left(u^{1-z} g\right)'(A_+)'+12 i  w \kappa  A_+ \theta '+\frac{1}{2} r_h^{z} u^{1-z} g (A_+)''=0
\ee
\be
\frac{u^{-1+z} (r_h)^{-z} w^2 A_-}{2g}+\frac{1}{2} r_h^{z}\left(u^{1-z} g\right)'(A_-)'-12 i  w \kappa  A_- \theta '+\frac{1}{2} r_h^{z} u^{1-z} g (A_-)''=0
\ee
The boundary behaviour of the fields at $u =0$ is
\be
A_\pm = A^0_{\pm} + A^1_{\pm}\frac{u^z}{z}+.. 
\ee
We seek solutions of these equations with infalling boundary conditions at the horizon of the form
\be
A_\pm \approx c_{\pm}\left(1-u\right)^{-\frac{i w}{(2z+1)r_h^z}}\left(1+A^{h}_{\pm}(1-u)+....\right),
\ee
where $c_{\pm}$ are constants. Then, as shown in \cite{Aprile:2010yb}, the combinations describing the components of the conductivity matrix in the holographic dictionary are
\be
\sigma_{yy}=\sigma_{xx} = \frac{r_h^z}{2iw }\left(\frac{A^1_+}{A^0_+}+\frac{A^1_-}{A^0_-}\right), \quad \sigma_{xy} =-\sigma_{yx}= \frac{r_h^z}{2 w}\left(\frac{A^1_+}{A^0_+}-\frac{A^1_-}{A^0_-}\right).
\ee

\begin{figure}
\centering
\begin{subfigure}{.5\textwidth}
  \centering
    \includegraphics[width=.9\linewidth]{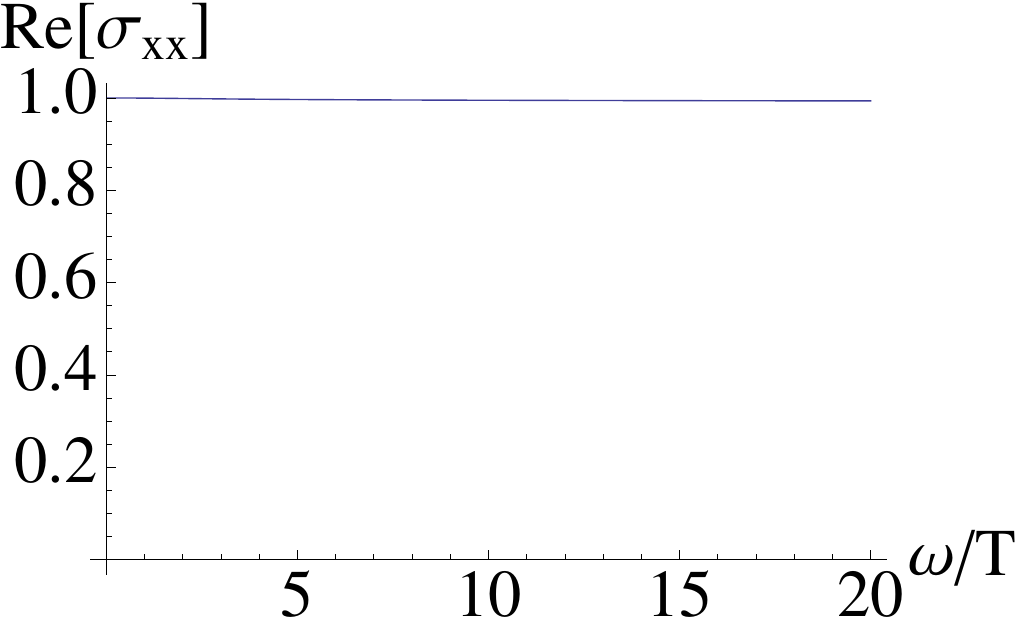}
  \caption{}
  \label{fig4a}
\end{subfigure}%
\begin{subfigure}{.5\textwidth}
  \centering
  \includegraphics[width=.9\linewidth]{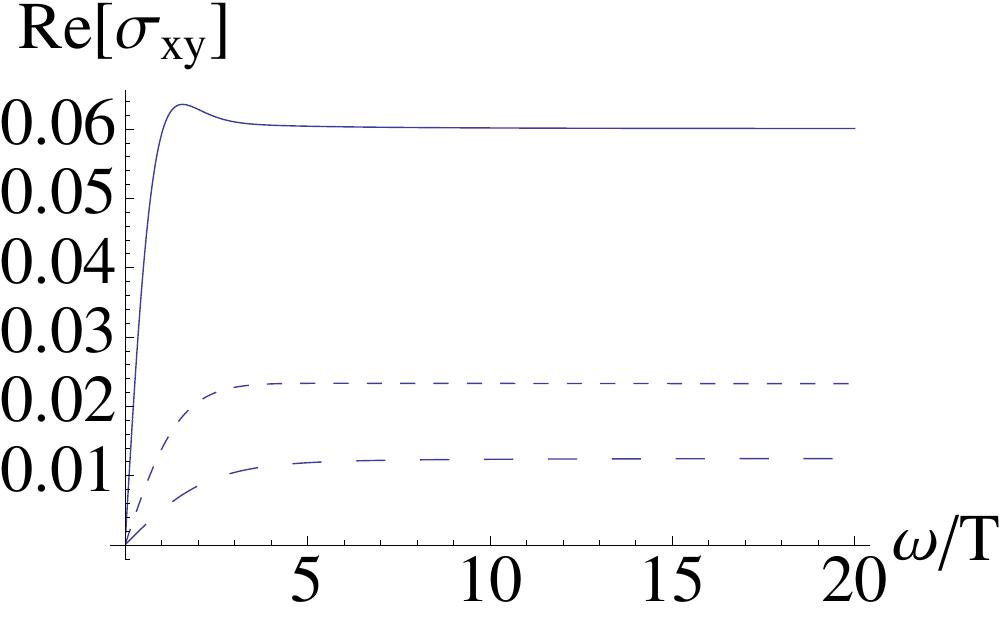}
  \caption{}
  \label{fig4b}
\end{subfigure}
\begin{subfigure}{.5\textwidth}
  \centering
  \includegraphics[width=.9\linewidth]{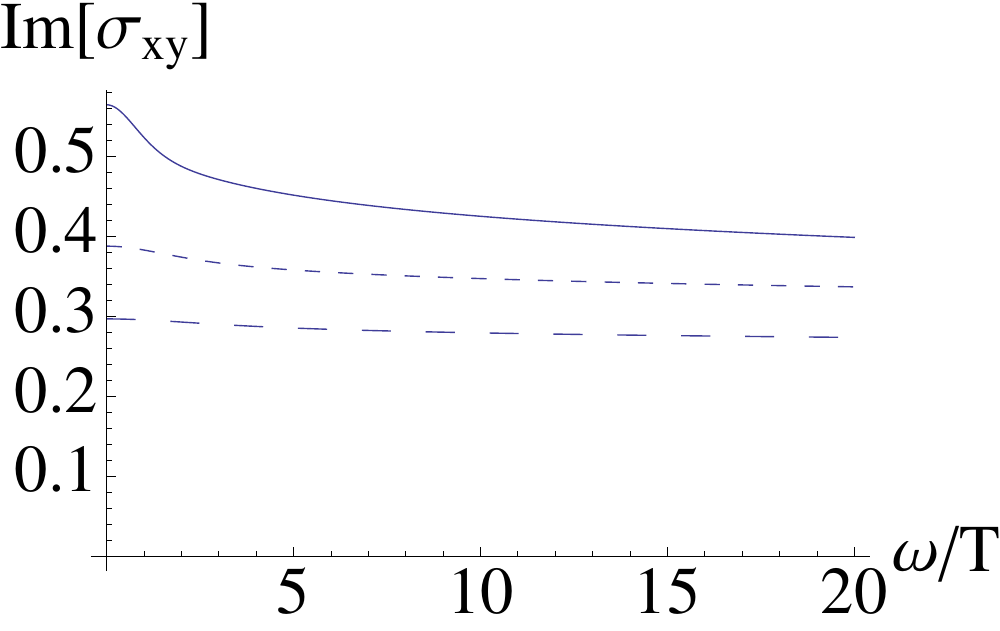}
  \caption{}
  \label{fig4b}
\end{subfigure}
\\
\caption{Conductivity of the normal phase. The plots correspond to $\kappa=0.5$, $m=0.1$ and $z=1,2,3$ for solid, thin dashed and large dashed profiles respectively.}
\label{hallcond}
\end{figure}

The numerical results are shown in Figure (\ref{hallcond}), we see that the diagonal components of the conductivity matrix respect the result of the relativistic scaling therefore we still observe that $\sigma_{xx}=\sigma_{yy}=1$ independently of $z$. As we expected a non vanishing axion coupling introduces non-diagonal components of the conductivity matrix. These components are small compared to the normal channel of charge transport. We see that $\sigma_{xy}$ tend to constants at large frequencies and rapidly go to zero as the frequency becomes small. We interpret this as a minimum energy of the perturbation modes required to excite the ``Hall" modes of conductivity. The constancy at large frequency is a consequence of the adimensionality of the conductivity in $2+1$ dimensions: as $\omega$ becomes large effectively the probe photon does not feel the finite temperature of the background and one loses all scales with which to make $\sigma_{ij}$ dimensionless (this is true for all $z$)\footnote{We thank S. Hartnoll for pointing this out to us.}.  As $z$ is increased the non-diagonal part of the conductivity becomes smaller which indicates that the axion effect is suppressed as one considers black holes with higher dynamical critical exponent. Thus, even though the axion field vanishes at the boundary we find a non-vanishing off-diagonal conductivity. A similar mechanism to observe such an effect was proposed in \cite{oai:arXiv.org:0704.1160} by having a constant axion field, in this case one would observe a constant off-diagonal conductivity at any temperature which differs from the above consideration. 

\section{Conclusions}

\quad  In this paper we investigated the combined effect of an axion coupling and non-relativistic Lifshitz scaling on finite temperature effects of holographic dual theories representing superconducting states. In particular we considered massive and massless axion fields. In order to observe the effects of the axion field a $\braket{p_x+ip_y}$ ansatz for the Yang-Mills field was used which, for the relativistic case with no axion, was known to be unstable \cite{Gubser:2008wv}. We found that below a critical temperature the Lifshitz black holes undergo a phase transition in which the gauge field condenses. It was shown that the free energy described by this phase is lower than the normal uncondensed phase below the critical temperature, signalling that a phase transition surely occurs. This was also confirmed by a stability analysis of the normal phase, demonstrating its instability below the critical temperature and its stability above it. We then demonstrated that the condensed phase is unstable, even in the case of a non-vanishing axion coupling or a non-relativistic scaling. It is most likely that the stable phase is represented by the $\braket{p_x}$ ansatz in which the axion field completely decouples from the system (even though no stability analysis for this phase exists). We further numerically computed the conductivity of the black holes in the normal phase and shown that the results of \cite{Herzog:2007ij} in which $\sigma_{xx} = \sigma_{yy}=1$ also hold for non-relativistic scaling. In this calculation we also showed that the axion field introduces non-diagonal ``Hall" components of the conductivity matrix. This conductivity is not superconducting, as it is computed in the normal phase with no condensates, and anomalous/topological in the sense that it is not generated by an external magnetic field (this kind of conductivity is known to exist, see for example \cite{Hall}).  We find that one needs a minimum perturbation energy to observe this type of transport and that it becomes constant at higher frequency. This effect is suppressed as one increases the Lifshitz scaling. \newline
 \indent An immediate important extension of this work concerns the backreaction of our probe system onto the geometry. In this sense one wishes to find full backreacting axion-Yang-Mills systems on Lifshitz black hole geometries, this has not yet been found to the extent of our knowledge. It is of crucial importance to the above stability calculations (and to the remaining as well) whether a back-reacting solution is still unstable. Let us point out that in a similar system a modification of the gravitational sector has been shown to give rise to stable axion phases \cite{Zayas:2011dw}, at least for relativistic scalings. In this context one may wish to extend this analysis to Lifshitz black holes and observe if a similar stabilization also occurs. One can also extend the above analysis to include other forms  for the axion potential (for example a linear $\lambda\theta$ term is also allowed as parity is already broken in this case) which might have important consequences on results investigated in this paper. Finally, one may also consider including fermions \cite{Li} in this system and investigate the consequence of their presence, in particular the existence of Fermi surfaces.

\section*{Acknowledgements}

The author would like to thank Tahsin Sisman, Fabrizio Canfora and Jorge Zanelli for useful discussions. This work was funded by Fondecyt grant no. 3140122. The Centro de Estudios Cient«õÞcos (CECS) is funded by the Chilean
Government through the Centers of Excellence Base Financing Program of Conicyt.

\end{document}